\documentstyle[aps,pre,manuscript]{revtex}   

\begin{document}

\draft
\title{Chaos and information entropy production}
\author{Bidhan Chandra Bag, Jyotipratim Ray Chaudhuri and 
Deb Shankar Ray{\footnote {e-mail : pcdsr@mahendra.iacs.res.in} } }
\address{Indian Association for the Cultivation of Science,
Jadavpur, Calcutta 700 032, INDIA.}
\maketitle
\begin{abstract}
We consider a general N-degree-of-freedom nonlinear 
system which is chaotic and dissipative and show that the nature 
of chaotic diffusion is reflected
in the correlation of fluctuation of linear stability matrix for the
equation of motion of the dynamical system whose phase space variables
behave as stochastic variables in the chaotic regime. Based on a 
Fokker-Planck description of the system in the associated tangent space
and an information entropy balance 
equation a relationship between chaotic diffusion and the 
thermodynamically-inspired quantities like entropy production and entropy
flux is established. The theoretical propositions have been verified by numerical
experiments.
\end{abstract}

\vspace{2.5cm}


\newpage

\section{Introduction}

Several authors have enquired recently about the relationship between phase 
space dynamics of a dynamical system and thermodynamics
\cite{kai,ono,pg,bonci,berdi,patt,ruelle,cohen,gaspard}. 
The question acquires 
a special relevance for the dissipative system when the phase space volume 
contracts by virtue of possessing the attractors and also when the system 
is nonlinear and comprises a few-degrees-of-freedom. Thus even when these
systems are not truly statistical in the thermodynamic sense,
it is possible that 
chaotic diffusion due to intrinsic deterministic chaos or stochasticity
plays a significant role in the dynamics. 
It is therefore
worthwhile to enquire about the relationship between {\it chaotic diffusion}
in a dynamical system and the {\it thermodynamically-inspired 
quantities like entropy production and entropy flux}.
Our purpose in this paper is to address this  specific issue.

In what follows we shall be concerned with 
the nonlinear dynamical systems which
are chaotic and dissipative. We do not consider any 
stochasticity due to thermal 
environment or external nonthermal noise. The ``deterministic 
stochasticity'' (i.e, chaos) has a purely dynamical basis and its emergence in nonlinear
dynamical system is essentially due to loss of correlation of initially
nearby trajectories. This is reflected in the linear stability
matrix or Jacobian of the system \cite{fox}. When chaos has fully set in, the time dependence
of this matrix can be described as a stochastic process, since the phase
space variables behave as stochastic variables \cite{casati}. It has been shown
that this fluctuation is amenable to a theoretical description in terms
of the theory of multiplicative noise \cite{van}. Based on this consideration a number
of important results of nonequilibrium statistical mechanics, like
Kubo relations, fluctuation-decoherence relation, fluctuation-dissipation
relation and exponential divergence of 
quantum fluctuations have been realized in chaotic dynamics of a
few-degree-of-freedom system \cite{sc1,sc2,sc3,bb1,bb2,bb3}. In the present paper we make use of this
stochastic description of chaotic dynamics to formulate a Fokker-Planck
equation of probability density function for the relevant dynamical
variables, the ``stochasticity'' (i.e, chaoticity) being incorporated through the fluctuations
of the time-dependent linear stability matrix.
Once the drift and chaotic diffusion terms are appropriately identified
the thermodynamic-like quantities can
be derived with the help of the suitable information entropy 
balance equation.

The outlay of the paper is as follows: In Sec. II we introduce a Fokker-Planck
description of the dynamical system and identify the chaotic 
drift and diffusion terms.
This is followed by setting up of an information entropy balance equation in
Sec. III. We then look for the entropy flux and entropy production-like terms
in the steady state. The 
shift of stationary state due to additional external forcing and the 
associated change in entropy production is considered in Sec. IV. We 
illustrate  the theory in detail with the help of an example in Sec. V. The
paper is concluded in Sec. VI. 

\section{A Fokker-Planck equation for dissipative 
chaotic dynamics}

We are concerned here with 
a general N-degree-of-freedom system whose Hamiltonian is
given by
\begin{equation}
H = \sum_{i=1}^N \frac{p_i^2}{2 m_i} +V(\{q_i\}, t) \; \;,
\; \; i = 1 \cdots N  \label{e1}
\end{equation}
where $\{q_i, p_i\}$ are the co-ordinate and 
momentum of the i-th degree-of-freedom, respectively, 
which satisfy the generic form of equations 

\begin{eqnarray}
\dot {q}_i = \frac{\partial H}{\partial p_i}  \; \;, \nonumber \\
\dot {p}_i = -\frac{\partial H}{\partial q_i}  \; \;.
\label{e2}
\end{eqnarray}

We now make the Hamiltonian system dissipative by introducing $-\gamma p_i$
on the right hand side of 
the second of Eqs.(2). 
For simplicity we assume $\gamma$ to be the
same for all the N degrees of freedom. 
By invoking the 
symplectic structure of the 
Hamiltonian dynamics as
\begin{eqnarray*}
z_i =
\left\{ \begin{array}{ll}
q_i & \; \; {\rm for} \; i=1 \cdots N  \; \;, \\
p_{i-N} & \; \; {\rm for} \; i=N+1, \cdots 2N \; \;,
\end{array}
\right.
\end{eqnarray*}
and defining I as 
\begin{eqnarray*}
I =
\left[
\begin{array}{cc}
0 & E \\
-E & -\gamma E
\end{array}
\right]
\end{eqnarray*}
where E is an $N\times N$ unit matrix, and $0$ is an  $N\times N$ null matrix,
the equation of motion for the dissipative system can be written 
as
\begin{equation}
\dot{z}_i = \sum_{j=1}^{2N} 
I_{ij} \frac{\partial H}{\partial z_j} \hspace{0.1cm}. \label{e3}
\end{equation}

We now consider two nearby trajectories, $z_i, \dot{z}_i$ and
$z_i+X_i$, $\dot{z}_i+ \dot{X}_i$ at the same time $t$
in $2N$ dimensional phase space. The time evolution of separation of these 
trajectories is then determined by
\begin{equation}
\dot{X}_i = \sum_{j=1}^{2N} J_{ij}(t) X_j \label{e4}
\end{equation}
in the tangent space or separation co-ordinate space ${X_i}$, where
\begin{eqnarray}
J_{ij}= \sum_k I_{ik} \frac{\partial^2 H}{\partial z_k \partial z_j}\; \; .
\label{e5}
\end{eqnarray}
Therefore the $2N \times 2N$ linear stability 
matrix $\underline{J}$ assumes the following form
\begin{equation}
\underline{J} =
\left[
\begin{array}{cc}
0 & E \\
M(t) & -\gamma E
\end{array}
\right]
\label{e6}
\end{equation}
where M is an $N\times N$ matrix. Note that the time dependence of 
stability matrix $\underline{J}(t)$ is due to the second derivative 
$\frac{\partial^2 H}{\partial z_k \partial z_j}$ which is determined by 
the equation of motion (3). 
The procedure for calculation of $X_i$
and the related quantities like Lyapunov exponents is to solve the 
trajectory equation (3) simultaneously
with Eq.(4). Thus when the dissipative 
system described by 
(3) is chaotic, \underline{J}(t) becomes 
(``deterministically'') stochastic phase space 
due to the fact that $z_i$-s behave as stochastic phase space variables
and the equation of motion (\ref{e4}) in the tangent space 
can be interpreted as a stochastic equation
\cite{sc1,sc2,sc3,bb1,bb2,bb3}.

In the next step we shall be concerned with a stochastic description of 
$\underline{J}(t)$ or $M(t)$. For convenience we split up M into two parts as
\begin{equation}
M = M_0 + M_1(t) \label{e7}
\end{equation}
where $M_0$ is independent of variables $\{z_i\}$ 
and therefore behaves a constant part
and $M_1$ is determined by the
variables $\{z_i\}$ for $i = 1 \cdots 2N$. 
$M_1$ refers to the fluctuating part.
We now rewrite the equation 
of motion (4) in the tangent space as
\begin{eqnarray}
\dot{X} & = & \underline{J} X  \nonumber \\
       & = & L\left(\left\{X_i\right\}, \left\{z_i\right\}\right)
       \label{e8}
\end{eqnarray}
where $X$ and $L$ are the vectors with $2N$ components. Corresponding to 
(\ref{e7}) $L$ in (\ref{e8}) can be split up again to yield

\begin{eqnarray}
\dot{X} & = & L^0(X) + L^1(X, \{z_i(t)\})   \nonumber \\
{\rm or} \; \; \; \; \; 
\dot {X}_i & = & {L_i}^0 (\{X_i\})
+L^1_i(\{X_i\},\{z_i\}) \; \; , \; \; \; \; i = 1 \cdots 2N \; \; .
\label{e9}
\end{eqnarray}

Eq.(\ref{e4}) indicates that Eq.(\ref{e8}) 
is linear in $\{X_i\}$. Eqs.(\ref{e4}), (\ref{e5}) and (\ref{e6})
express the fact the first $N$ components of $L^1$ are zero and the last $N$ 
components of $L^1$ are the functions of $\{X_i\}$ for $i = 1 \cdots N$ only.
The fluctuation in $L_i^1$ is caused by the chaotic variables 
$\left\{z_i\right\}$-s. 
By defining $\nabla_X$ as
differentiation with respect to components of $X$, i. e. , $\{X_i\}$ (explicitly 
$X_i = \Delta q_i$ for $i = 1 \cdots N$ and  $X_i = \Delta p_i$ for 
$i = N+1 \cdots 2N$) and since $L_i^1 = 0$ for $ i =1 \cdots N$ and 
$L_i^1 = L_i^1 (X_1 \cdots X_N)$ for $ i = N+1 \cdots 2N$
we have
$\nabla_X \cdot L^1 = \sum_{i=1}^N (\frac{\partial}{\partial X_i} \cdot 0)  +
\sum_{i=N+1}^{2N} \frac{\partial}{\partial X_i} \cdot L_i^1(X_1 \cdots X_N) = 0$.
This allows us to write the following relation (which will be used later on),
\begin{equation}
\nabla_X \cdot L^1 \phi(\{X_i\}) = L^1 \cdot \nabla_X \phi(\{X_i\}) \; \;
\label{e10}
\end{equation}
where $\phi(\{X_i\}$) is any function of $\{X_i\}$. 

Note that Eq.(\ref{e9}) by virtue of (\ref{e8}) 
is a linear differential equation
with multiplicative ``noise'' due to $\{z_i\}$ determined by 
equation of motion (\ref{e3}). 
This is the starting point of our further analysis.

Eq.(\ref{e9}) determines a stochastic process with 
some given initial conditions
$\{X_i(0)\}$. We now consider the motion of a representative point $X$ in
$2N$ dimensional tangent space ($X_1 \cdots X_{2N}$) as governed by 
Eq.(\ref{e9}).
The equation of continuity, which expresses the conservation of points
determines the variation of density function $\phi(X, t)$ in time as given by

\begin{equation}
\frac{\partial \phi(X, t)}{\partial t} = - {\bf \nabla_X} \cdot L(t) 
\phi(X, t) \; \;.
\label{e11}
\end{equation}
Expressing $A_0$ and $A_1$ as
\begin{eqnarray}
A_0 & = & - \nabla_X \cdot L^0  \nonumber \\
{\rm and} \; \; \; \; \; A_1 & = & - \nabla_X \cdot L^1 
\label{e12}
\end{eqnarray}
we may rewrite the equation of continuity as
\begin{equation}
\frac{\partial \phi(X, t)}{\partial t} = [A_0 + \alpha A_1(t)] 
\phi(X, t) \; \;.
\label{e13}
\end{equation}
It is easy to recognize that while $A_0$ denotes the constant part $A_1$ contains the 
multiplicative fluctuations through the phase space variables of the
dynamical system $\{z_i(t)\}$. $\alpha$ is a parameter
introduced from outside to keep track of the order of fluctuations in the 
calculations. At the end we put $\alpha =1$. 

One of the main results for the linear equations of the form 
(\ref{e13}) with 
multiplicative noise may now be in order \cite{van}. The average equation of 
$\langle \phi \rangle$ obeys [ $P(X, t) \equiv \langle \phi \rangle$],
\begin{equation}
{\dot P} = \left \{ A_0 + \alpha \langle A_1 \rangle
+ \alpha^2 \int_0^\infty d\tau
\langle \langle A_1(t) \exp (\tau A_0) A_1(t-\tau) 
\rangle \rangle \exp (-\tau A_0) \right \} P(X, t) \; \; .
\label{e14}
\end{equation}

The above result is based on second order cumulant expansion and is 
valid when fluctuations are small but rapid and the correlation 
time $\tau_c$ is short but finite or more precisely

\begin{equation}
\langle \langle A_1(t) A_1(t') \rangle \rangle = 0 \; \; 
\; \; \; {\rm for} \; \; 
\left|t-t'\right| \; > \; \tau_c \;.
\label{e15}
\end{equation}
We have, in general, $\langle A_1 \rangle \ne 0$. Here  
$\langle \langle \cdots \rangle \rangle$ implies
$\langle \langle \zeta_i \zeta_j \rangle \rangle 
= \langle \zeta_i \zeta_j \rangle -\langle \zeta_i \rangle \langle \zeta_j 
\rangle $.

The Eq.(\ref{e14}) is exact in that limit $\tau_c \rightarrow 0$. 
Making use of relation (\ref{e12}) in (\ref{e11}) we obtain 
\begin{eqnarray}
\frac{\partial P}{\partial t} & = & 
\left\{ -\nabla_X \cdot {\bf L}^0  -\alpha \langle \nabla_X \cdot 
{\bf L}^1 \rangle + \alpha^2 \int_0^\infty d\tau
\langle \langle \nabla_X \cdot{\bf L}^1(t)   \exp ( -\tau \nabla_X \cdot{\bf L}^0)  
\right.\nonumber\\
& & \left. \nabla_X \cdot {\bf L}^1(t-\tau) \rangle \rangle
\exp ( \tau \nabla_X \cdot {\bf L}^0 )  \right\} P \; \; .
\label{e16}
\end{eqnarray}
The above equation can be transformed into the following Fokker-Planck
equation ($\alpha = 1$) for probability density function $P(X,t)$, 
(the details are given in the Appendix A);
\begin{equation}
\frac{\partial P(X,t)}{\partial t} = -\nabla_X . F P +
\sum_{i,j} {\cal D}_{ij} \frac{\partial^2 P}{\partial X_i \partial X_j}
\label{e17}
\end{equation}
where,
\begin{equation}
F=L^0 + \langle L^1 \rangle + Q
\label{e18}
\end{equation}
and $Q$ is a $2N$-dimensional vector whose components are
defined by
\begin{eqnarray}
Q_j=-\int_0^\infty \langle \langle R'_j \rangle \rangle
d \tau det_1(\tau) det_2(\tau) \; \; \;.
\label{e19}
\end{eqnarray}
Here the determinants $det_1$, $det_2$ and $R'_j$ are given by
\begin{eqnarray}
det_1(\tau)& = & \left|\frac{d X^{-\tau}}{d X} \right|
\; \; \; \; {\rm and} \; \; det_2(\tau)=
\left|\frac{d X}{d X^{-\tau}} \right|
\nonumber \\
{\rm and} \; \; \; R'_j & = & \sum_i L_i^1(X,t) \frac{\partial}{\partial X_i}
\sum_k L_k^1(X^{-\tau},t-\tau) \frac{\partial X_j}{\partial X_k^{-\tau}}
\; \; .
\label{a20}
\end{eqnarray}

It is easy to recognize $F$ as an evolution operator. Because 
of the dissipative perturbation we note that div $F < 0$.

The diffusion coefficient ${\cal D}_{ij}$ in Eq.(\ref{e17}) is defined as
\begin{equation}
{\cal D}_{ij}=\int_0^\infty \sum_k \langle \langle L_i^1(X,t) 
L_k^1(X^{-\tau},t-\tau)\frac{dX_j}{dX_k^{-\tau}}
\rangle \rangle d\tau \; \; \;.
\label{e21}
\end{equation}
We have followed closely van Kampen's approach \cite{van} to generalized Fokker-Planck
equation (\ref{e17}). Before concluding this section several critical
remarks regarding this derivation need attention:

First, the process $M_1(t)$ determined by $\{z_i\}$ is
obtained {\it exactly} by solving equations of motion (\ref{e3}) for the
chaotic motion of the system. It is therefore necessary to emphasize that we have 
{\it not assumed} any special property of noise, such as, $M_1(t)$
is Gaussian or $\delta$-correlated. We reiterate Van Kampen's
emphasis in this approach.

Second, the only assumption made about the noise is that its
correlation time $\tau_c$ is short but finite compared to the
coarse-grained timescale over which the average quantities evolve.
Or in other words the velocity changes should be small, smooth and 
uncorreleted after short times. This assumption, however, puts a restriction
on the applicability of the present theory to certain class of systems, e. g. ,
systems subjected to `hard' collisions such as billiards and molecular
systems also in certain non-dissipative Hamiltonian systems like standard map
for which the usual assumptions about the rapid decay of correlations and
fluctuations are not valid and entropy production does not occur. The special 
reference may be made in this connection to the work of Zaslavsky and
collaborators \cite{zas} to demonstrate that in many real systems 
the decay of correlation exhibits power law dependence, 
distributions admit of infinite moments and the fluctuations become 
long-lasting. Since the mathematical difficulties of dealing with finite 
arbitrary correlation time of noise in a chaotic system is quite formidable we
confine ourselves in the present discussion to chaotic systems with short
but finite noise correlation time.

Third, we take care of fluctuations upto second order which implies 
that the deterministic noise is not too strong.

Eq.(\ref{e17}) is the required Fokker-Planck equation in the tangent space
$\{X_i\}$. However the important point is to note 
that the drift and diffusion terms are
determined by the phase space $\{z_i\}$ properties of the chaotic system and
directly depend on the correlation function of the fluctuations of the
second derivatives of the Hamiltonian (\ref{e5}).

\section{Information entropy balance: Entropy production}

We shall now consider the well-known relation between probability 
density function $P(X,t)$ and information entropy $S$ as given by

\begin{equation}
S = -\int dX P(X,t) \ln P(X,t) \; \;.
\label{e22}
\end{equation}

Note that in the above definition of entropy we use $P(X, t)$, the probability
distribution function in the tangent space since
one is concerned here with the
expansion of the phase space in terms of a tangent space and dialation 
coefficients of the dynamical system for which the expanding and
contracting manifolds can be defined. On the other hand it is worthwhile
to recall the dynamic 
entropy of a dynamical system (Kolmogorov entropy) defined in terms of the properties
of evolution in the tangent space. A remark on the connection between
entropy and expansion by Sinai \cite{sinai} is noteworthy in this context ; `` It already 
seems clear that positiveness of the entropy and presence of mixing
is related to extreme instability of the motion of the system : 
trajectories emanating from the nearby points must, generally speaking,
diverge with exponential velocity. Thus entropy 
is characterized here by the speed of approach of the asymptotic trajectories''
which is formalized by defining the expansion coefficient as logarithm
of the relative increase under the flow of a volume element in the 
expanding manifold. Our definition
of information entropy (22) makes use of the tangent space description 
of the systems in terms of logarithm of the
probability of expansion in the tangent space, keeping in mind that
$-\ln P$ is ``a measure of unexpectedness of an event (the amount of information)
and the information entropy is a mean value of this unexpectedness for the 
entire system'' \cite{Vasi}. The definition (22) is therefore different
from Kolmogorov entropy. 
We emphasize that
even in absence of any direct formal connection between $P(X, t)$ and the
phase space distribution function it is possible to use the distribution function $P(X, t)$ defined 
in the tangent space to have an explicit expression for entropy
production-like quantity as a function of the properties of phase space
variables $\{z_i\}$ of the dynamical system (i. e. , in terms of drift and 
diffusion coefficients of the Fokker-Planck equation).

The above definition of an information entropy like quantity
allows us to have an evolution equation for entropy.
To this end we observe from Eqs.(\ref{e17}) and (\ref{e22}) that 
\cite{nicolis,jou}

\begin{equation}
\frac{dS}{dt} = -\int dX \left[ - \sum_i \frac{\partial}{\partial X_i}(F_i P)
+ \sum_i \sum_j \int {\cal D}_{ij}
\frac{\partial^2 P}{\partial X_i \partial X_j}
\right] \ln P \; \; \;.
\label{e23}
\end{equation}

Note that the probability density function P(X,t) is defined in tangent
space $\{X_i\}$. ${\cal D}$ and $F$ as expressed in 
Eqs.(\ref{e21}) and (\ref{e18}), respectively,
are determined by the correlation functions of fluctuations of
second derivative of the Hamiltonian of the system. 
Eq.(\ref{e23}) therefore suggests 
that the
entropy production-like term originating from 
Eq.(\ref{e23}) is likely to bear the signature
of the chaotic dynamics. The relation is direct and general as it is evident
from the following equation [obtained after partial 
integration of Eq.(\ref{e23}) with the 
natural boundary condition on $P(X,t)$ that it vanishes as 
$|X|\rightarrow\infty$ and assuming the $X$-dependence of ${\cal D}_{ij}$ to 
be weak (as a first approximation)];

\begin{equation}
\frac{dS}{dt} = \int dX P \nabla_X \cdot F
+ \sum_i \sum_j {\cal D}_{ij} \int \frac{1}{P} 
\frac{\partial P}{\partial X_i}\frac{\partial P}{\partial X_j} dX \; \; \;.
\label{e24}
\end{equation}

The first term in (\ref{e24}) has no definite sign while 
the second term is positive 
definite because of positive definiteness of ${\cal D}_{ij}$. Therefore the 
second one can be identified \cite{nicolis} as the entropy production
\begin{equation}
S_{prod} = \sum_i \sum_j {\cal D}_{ij} \int \frac{1}{P_s} 
\frac{\partial P_s}{\partial X_i}\frac{\partial P_s}{\partial X_j} dX
\label{e25}
\end{equation}
in the steady state. The subscript $s$ 
of $P_s$ refers to steady state. It is evident from Eq.(\ref{e24}) that

\begin{eqnarray}
S_{flux} & = & \int dX \; P_s(X) \; \nabla_X \cdot F \; \; 
 =  {\overline{\nabla_X \cdot F} } \nonumber\\
S_{prod} & = & -S_{flux} \; \;.
\label{e26}
\end{eqnarray}

\noindent
Note that since the chaotic system is dissipative
$\overline{\nabla_X . F}$ is negative [See Eq.(\ref{e18})].

It is thus evident that the relations 
(\ref{e25}), (\ref{e26})  illustrate the dynamical origin of 
entropy production like quantity 
in a chaotic dissipative system. The dynamical 
signature is manifested through 
the drift term $F$ and the chaotic diffusion terms in ${\cal D}_{ij}$.
It must be emphasized that the notion of diffusion has nothing to do
with any external reservoir. Rather it pertains to intrinsic
diffusion in phase space of the chaotic system itself.

\section{The chaotic system driven by an external force} 

We shall now examine the entropy production when the dissipative
chaotic system is thrown away from the steady state due to an
additional weak applied force. To this end we consider the drift $F_1$ due to 
external force 
so that the total drift $F$ has now two contributions

\begin{equation}
F(X) = F_0(X) + h F_1(X) \; \;.
\label{e27}
\end{equation}

When $h = 0$, $P = P_s$. The deviation of $P$ from $P_s$ 
in presence of nonzero
small $h$ can be explicitly taken into account once we make use of the identity
for the diffusion term \cite{nicolis}

\begin{equation}
\frac{\partial^2 P}{\partial X_i \partial X_j} = \frac{\partial}{\partial X_i}
\left[ P \frac{\partial \ln P_s}{\partial X_j}\right] +
\frac{\partial}{\partial X_i}\left[ P_s \frac{\partial}{\partial X_j}\frac{P}{P_s}\right] 
\; \; .
\label{e28}
\end{equation}

When $P = P_s$ the second term in 
(\ref{e28}) vanishes. In presence of additional
forcing the Eq.(\ref{e17}) becomes,
\begin{equation}
\frac{\partial P}{\partial t} = - \nabla_X . \psi P - h \nabla_X . F_1 P
+ \sum_i \sum_j {\cal D}_{ij} \frac{\partial}{\partial X_i} \left( P_s \frac
{\partial}{\partial X_j}\frac{P}{P_s} \right)
\label{e29}
\end{equation}
where $\psi$ is defined as
\begin{equation}
\psi = F_{0} - \sum_j {\cal D}_{ij} \frac{\partial \ln P_s}{\partial X_j} 
\; \; .
\label{e30}
\end{equation}
Here we have assumed for simplicity that ${\cal D}_{ij}$ is not
affected by the additional forcing. The leading order influence
is taken into account by the additional drift term in Eq.(\ref{e29}).

Under steady state condition $(P = P_s)$ and $h = 0$, the second and 
the third 
terms in (\ref{e29}) vanish yielding

\begin{equation}
\nabla_X . \psi P_s = 0 \; \;.
\label{e31}
\end{equation}
It is immediately apparent that $\psi P_s$ refers to a current ${\cal J}$ 
where ${\cal J} = \psi P_s$. The steady state condition therefore reduces to
an equilibrium condition (${\cal J}=0$) if
\begin{equation}
\psi = 0 \; \; .
\label{e32}
\end{equation}
(In Sec. V we shall consider an explicit example to show that $\psi=0$).
This suggests a formal relation between $F_{0i}$ and ${\cal D}_{ij}$ as
\begin{equation}
F_{0i} = \sum_j {\cal D}_{ij} \frac{\partial \ln P_s}{\partial X_j}
\label{e33}
\end{equation}
where $P_s$ may now be referred to as the {\it equilibrium} density function
in separation co-ordinate space. 
$F_0$  contains dissipation constant $\gamma$
and the diffusion matrix 
${\cal D}_{ij}$ is a function of correlation function of 
fluctuations of the second derivative of the Hamiltonian. 

To consider the information entropy 
balance equation in presence of external forcing
we first differentiate 
Eq.(\ref{e22}) with respect to time and use Eq.(\ref{e29}) to 
obtain
\begin{eqnarray}
\frac{d S}{d t} & = & -\frac{d}{d t} \int dX P \ln P_s \nonumber \\
& & - \int dX \ln{\frac{P}{P_s}}
\left[ -\nabla_X \cdot \psi P 
-  h \nabla_X \cdot F_1 P +
\sum_i \sum_j {\cal D}_{ij} \frac{\partial}{\partial X_i} \left( P_s \frac
{\partial}{\partial X_j}\frac{P}{P_s} \right)\right] \; \;.
\label{e34}
\end{eqnarray}

It is apparent that as $P$ deviates 
from $P_s$, $\frac{P}{P_s}$ differs from unity
and the entire second integral within the parenthesis \cite{nicolis}
is of the second order. Note that 
$\ln P_s$  in the first integral in Eq.(\ref{e34})
is a constant of motion and the integral denotes its average. 
The first term vanishes because it is of higher order as it involves $P^2$
and others. (Moreover since in the discussion that follows we consider 
the steady state, this term does not contribute to the subsequent calculations). 
To compute the contribution $\Delta S$ to entropy balance due to the
external forcing only we perform integration of 
the second, third and fourth terms
by parts. We thus obtain,

\begin{eqnarray}
\frac{d \Delta S}{d t} & = & 
h^2 \int dX \delta P \nabla_X \cdot F_1 + h^2 \int
dX \left(\sum_i F_{1i} \frac{\partial \ln P_s}{\partial X_i} 
\right) \delta P  \nonumber \\
& + & \sum_i \sum_j {\cal D}_{ij} \int dX P 
(\frac{\partial}{\partial X_i} \ln{\frac{P}{P_s}}) 
(\frac{\partial}{\partial X_j} \ln{\frac{P}{P_s}}) 
\label{e35}
\end{eqnarray}
where we have put $h \delta P = P-P_s$. In the new steady state
(in presence of $h \ne 0$), the entropy production and the flux like
terms balance each other as follows:
\begin{equation}
\Delta S_{prod} = - \Delta S_{flux}
\label{e36}
\end{equation}
with
\begin{equation}
\Delta S_{prod} = \sum_{i,j} {\cal D}_{ij} \int dX P\left(
\frac{\partial}{\partial X_i}
\ln \frac{P}{P_S}\right) \left(\frac{\partial }{\partial X_j} 
\ln \frac{P}{P_S}\right)
\label{e37}
\end{equation}
and
\begin{equation}
\Delta S_{flux} = h^2 \int dX \delta P \nabla_X . F_1 +h^2
\int dX \left( \sum_i F_{1i} \frac{\partial \ln P_S}{\partial X_i}\right)
\delta P \; \; \;.
\label{e38}
\end{equation}

In the following section we shall work out a specific example to provide
explicit expressions for the entropy production 
and some related quantities due to external forcing.

\section{Applications}

\subsection{Entropy production in the steady state}

To illustrate the theory developed above, we now choose a driven
double-well oscillator system with Hamiltonian
\begin{equation}
H=\frac{p_1^2}{2}+aq_1^4-bq_1^2+ \epsilon q_1\cos\Omega t
\label{e39}
\end{equation}
where $p_1$ and $q_1$ are the momentum and position variables of the 
system. $a$ and $b$ are the constants characterizing the potential.
$\epsilon$ includes the effect of coupling constant and the
driving strength of the external field with frequency $\Omega$. This model
(\ref{e39}) has been extensively used in recent years for the study of chaotic 
dynamics \cite{sc1,sc2,sc3,lin}.

The dissipative equations of motion for the tangent space variables $X_1$ 
and $X_2$ corresponding to $q_1$ and $p_1$ (Eq.8) read as follows:
\begin{equation}
\frac{d}{d t} \left[
\begin{array}{c}
X_1 \\
X_2
\end{array} \right]
= \underline{J} \left[
\begin{array}{c}
X_1 \\
X_2
\end{array} \right] \; \; , \; \;
\left \{
\begin{array}{c}
\Delta q_1 = X_1 \\
\Delta p_1 = X_2
\end{array} \right \} \; \; ,
\label{e40}
\end{equation}
where $\underline{J}$ as expressed in our earlier notation
\begin{eqnarray*}
z_1 = q_1 \; \; \; \; \; 
z_2 = p_1
\end{eqnarray*}
is given by
\begin{eqnarray*}
\underline{J} = \left(
\begin{array}{cc}
0 & E \\
-E & -\gamma E
\end{array}
\right)
\left(
\begin{array}{cc}
\frac{\partial^2 H}{\partial z_1 \partial z_1} &
\frac{\partial^2 H}{\partial z_1 \partial z_2} \\
\frac{\partial^2 H}{\partial z_2 \partial z_1} & 
\frac{\partial^2 H}{\partial z_2 \partial z_2}
\end{array}
\right) \; \; .
\end{eqnarray*}
Therefore $\underline{J}$ reduces to
\begin{eqnarray*}
\left(
\begin{array}{cc}
0 & 1 \\
\zeta (t)+2b \;  & \; -\gamma
\end{array}
\right) \; \; ,
\end{eqnarray*}
where $\zeta(t)=-12 a z_1^2$. Thus we have
\begin{eqnarray*}
M_0 = 2b \; \; \; \; M_1 = \zeta(t) \; \; .
\end{eqnarray*}
Eq.(\ref{e40}) is thus rewritten as 
\begin{equation}
\frac{d}{d t} \left(
\begin{array}{c}
X_1 \\
X_2
\end{array}
\right) = L^0+L^1
\label{e41}
\end{equation}
with
\begin{eqnarray*}
L^0=\left(
\begin{array}{c}
X_2\\
2bX_1-\gamma X_2
\end{array}
\right)
\; \; \; \; {\rm and} \; \;
L^1=\left(
\begin{array}{c}
0\\
\zeta (t) X_1
\end{array}
\right) \; \; ,
\end{eqnarray*}
where $L^0$ and $L^1$ are the constant and the 
fluctuating parts, respectively.
The fluctuations in $L^1$, i.e., in $\zeta(t)$ is due to stochasticity of the
following chaotic dissipative dynamical equations of motion;
\begin{eqnarray}
\dot{z}_1 & = & z_2 \nonumber \\
\dot{z}_2 & = & -az_1^3+2bz_1-\epsilon \cos\Omega t -\gamma z_2 \; \; .
\label{42}
\end{eqnarray}
Now for the constant part and the fluctuating part we write
\begin{eqnarray*}
L^{01}= X_2 \; \; \; \; &&L^{02} = 2bX_1-\gamma X_2 \\
L^{11}= 0 \; \; \; \; &&L^{12} = \zeta (t) X_1 \; \; \;.
\end{eqnarray*}
We may then apply the result of Eq.(\ref{a5}).

The mapping $X \rightarrow X^t$ is found by solving the `unperturbed'
equations
\begin{eqnarray*}
\dot{X}_1 & = & X_2 \\
\dot{X}_2 & = & G_2-\gamma X_2
\end{eqnarray*}
Comparison with Eq.(\ref{a7}) shows that $G_2(= 2bX_1)$ is free from $X_2$.

As a short time approximation we consider the variation of $X_1$ and 
$X_2$ during $\tau_c$ ;
\begin{eqnarray}
X_1^{-\tau} & = & -\tau X_2+X_1 = \bar{G}_1(X_1,X_2) \nonumber \\
X_2^{-\tau} & = & -\tau G_2+e^{\gamma \tau} X_2 
= \bar{G}_2(X_1,X_2) \; \; \;.
\end{eqnarray}
So the $g$-matrix of Eq.(\ref{a15}) becomes
\begin{equation}
\underline{g} = \left(
\begin{array}{cc}
1 & \tau e^{-\gamma \tau} \\
2b\tau e^{-\gamma \tau} & e^{-\gamma \tau} \; \;
\end{array}
\right) \; \;.
\end{equation}
The vector $R$ from Eq.(\ref{a15}) can then be identified as

\begin{eqnarray}
R & = & \left(
\begin{array}{c}
\zeta(t-\tau)(X_1-\tau X_2) g_{12}\\
\zeta(t-\tau)(X_1-\tau X_2) g_{22}
\end{array}
\right) \nonumber \\
& = & \left(
\begin{array}{c}
\zeta(t-\tau)X_1 \tau e^{-\gamma \tau}\\
\zeta(t-\tau)(X_1-\tau X_2) e^{-\gamma \tau}
\end{array}
\right) \; . \; \; \; 
{\rm [\; neglecting \; the \; terms \; of \; 0(\tau^2) \;]} \; \; .
\end{eqnarray}

Similarly the vector $R'$ is given by
\begin{equation}
R' = \left(
\begin{array}{c}
0\\
-\zeta(t-\tau)\tau e^{-\gamma \tau} X_1 \zeta(t)
\end{array}
\right) \; \; .
\end{equation}
From Eqs.(43) and (44) we have
\begin{equation}
det_1(\tau) det_2(\tau) \simeq 1   \; \; .
\end{equation}
Then the vector $Q$ can be written as
\begin{equation}
Q = \left(
\begin{array}{c}
0 \\
X_1 \int_0^\infty \langle \langle \zeta(t) \zeta(t-\tau) \rangle \rangle
\tau e^{-\gamma \tau} d \tau
\end{array}
\right) \; \; .
\end{equation}
Now the diffusion matrix ${\cal D}$ can be constructed as
\begin{equation}
{\cal D} = \left(
\begin{array}{cc}
0 & 0 \\
{\cal D}_{21} & {\cal D}_{22} 
\end{array}
\right)
\end{equation}
where 
\begin{eqnarray*}
{\cal D}_{21}=
X_1^2(0)\int_0^\infty \langle\langle \zeta(t) \zeta(t-\tau) \rangle \rangle
\tau e^{-\gamma \tau}d\tau
\end{eqnarray*}
and
\begin{eqnarray*}
{\cal D}_{22}=
X_1^2(0)\int_0^\infty \langle\langle \zeta(t) \zeta(t-\tau) \rangle \rangle
e^{-\gamma \tau}d\tau 
- X_1(0)X_2(0)\int_0^\infty \langle\langle \zeta(t) \zeta(t-\tau) 
\rangle \rangle
\tau e^{-\gamma \tau}d\tau  \; \; .
\end{eqnarray*}

It is important to mention that the assumption of weak $X$-dependence of the
diffusion coefficient (by freezing its time dependence) is permitted as a first approximation within the perview
of the present second order theory for which the strength of noise is not too
large. We also emphasize that for actual theoretical estimate of the entropy
production in terms of the formulae (26) or (58) explicit evaluation of the
diffusion coefficients is not required (see the next section). A straightforward
calculation of drift is sufficient for the purpose. This point will be
clarified in greater detail in the next Sec.VB.

Then the Fokker-Planck equation (17) for the dissipative driven double-well
oscillator assumes the following form:
\begin{eqnarray}
\frac{\partial P}{\partial t} & = & -X_2 \frac{\partial P}{\partial X_1}
-\omega^2 X_1 \frac{\partial P}{\partial X_2} + \gamma 
\frac{\partial}{\partial X_2} (X_2 P) +{\cal D}_{21} \frac{\partial^2 P}
{\partial X_2 \partial X_1} + {\cal D}_{22} \frac{\partial^2 P}
{\partial X_2^2}
\end{eqnarray}
where 
\begin{eqnarray*}
\omega^2 = 2b+c+c_2
\end{eqnarray*}
with
\begin{eqnarray}
c_2 & = & \int_0^\infty \langle\langle \zeta(t)\zeta(t-\tau) \rangle \rangle 
\tau e^{-\gamma \tau} d\tau \nonumber \\
c & = & \langle \zeta \rangle \; \; .
\end{eqnarray}

\noindent
The similarity of the equation (50) to generalized Kramers' equation can not
be overlooked. This suggests a clear interplay of chaotic diffusive motion
and dissipation in the dynamics.

We now let
\begin{equation}
U=a_s X_1+X_2
\end{equation}
where $a_s$ is a constant to be determined.

Then under steady state condition Eq.(50) reduces to the following form :
\begin{equation}
\frac{\partial}{\partial U}(\lambda_s U)P_s + {\cal D}_s 
\frac{\partial^2 P_s}{\partial U^2} = 0  \; \; ,
\end{equation}

where 
\begin{eqnarray*}
{\cal D}_s = {\cal D}_{22} + a_s {\cal D}_{21} \; \; ,
\end{eqnarray*}
and
\begin{eqnarray*}
\lambda_s U = -a_sX_2-\omega^2 X_1+\gamma X_2   \; \; .
\end{eqnarray*}
Here $\lambda_s$ is again a constant to be determined. Putting (52) in
$\lambda_s U$ as given above and comparing the coefficients of $X_1$ and $X_2$ we obtain
\begin{eqnarray*}
\lambda_s a_s = -\omega^2 \; \; \; \; 
{\rm and} \; \; \lambda_s = -a_s+\gamma  \; \;.
\end{eqnarray*}
The physically allowed solutions for $a_s$ and $\lambda_s$ are as follows ;
\begin{equation}
a_s = \frac{\gamma - \sqrt{\gamma^2+4\omega^2}}{2}
\; \; \; \; \; {\rm and} \; \;
\lambda_s=\frac{\gamma+\sqrt{\gamma^2+4\omega^2}}{2} \; \; .
\end{equation}
The stationary solution of (53) $P_s$ is given by
\begin{equation}
P_s = N e^{-\frac{\lambda_s U^2}{2{\cal D}_s}} \; \;.
\end{equation}
Here $N$ is the normalization constant. By virtue of (55) $\psi$ 
corresponding to Eq.(30) is therefore
\begin{equation}
\psi = \lambda_s U - {\cal D}_s \frac{\partial \ln P_s}{\partial U} = 0 \; \;.
\end{equation}
Since $\psi P_s$ defines a current, $P_s$
defines a zero current situation or an equilibrium condition.

The equilibrium solution $P_s$ from (55) can now be used to calculate the
steady state entropy production as given by Eq.(25). We thus have
\begin{equation}
S_{prod}={\cal D}_s\int_{-\infty}^\infty  \frac{1}{P_s}
\left(\frac{\partial P_s}{\partial U}\right)^2 dU \; \; .
\end{equation}
Explicit evaluation shows
\begin{equation}
S_{prod} = \lambda_s \; \; ,
\end{equation}
where $\lambda_s$ is given by Eq.(54).

The above result demonstrates a rather straightforward connection
between entropy production-like quantity of a chaotic system and 
the dynamics through
dissipation constant $\gamma$, parameters of the Hamiltonian and
correlation of fluctuations of the second derivatives of the Hamiltonian 
in the steady state. It is important to note that since one is concerned
here with a few degree-of-freedom system with no explicit reservoir 
temperature does not appear in the expression for entropy production like
term (58). The entropy production in a truly thermodynamic system and in the
present case are therefore distinct.

\subsection{Numerical Verifications}
To verify the above theoretical analysis in terms of numerical experiments
we now concentrate on the following two points. First it is necessary to
establish numerically that the dynamical system reaches a steady state, i . e .,
the probability density function $P(\{X_i\}, t)$ attains a steady state
distribution $P_s(\{X_i\})$ in the long time limit. Second, the entropy production
in the steady state calculated by the formula (58) needs to be verified
numerically. To address the first issue we now proceed as follows ;

The dissipative chaotic dynamics corresponding to the model Hamiltonian (39) is
governed by Eqs.(40) and (42). We choose the following values of the 
parameters \cite{lin} $ a = 0.5$, $b = 10$, $\omega = 6.07$ and $\gamma = 0.001$.
The coupling-cum-field strength $\epsilon$ is varied from set to set.
We fix the initial condition $z_1(0) = -3.5$, $z_2(0) = 0$ which ensures strong global
chaos \cite{lin}. To determine the steady state distribution function, say,
$P_s(X_1)$, where $X_1 = \Delta q$ (Eq. 40) from the dynamical point of view
we first define $d_0$ as the separation of the two initially nearby trajectories
and $d(t)$ as the corresponding separation at time t. To express $d(t)$ 
we write $d(t) = [\sum_i^N(X_i)^2 +\sum_{i = N+1}^{2N}(X_i)^2]^{\frac{1}{2}}$.
$d(t)$ is determined by solving numerically Eqs.(40) and (42), simultaneously
for the initial conditions of $z_1$ and $z_2$ corresponding to Eq.(42). To
follow the evolution of $X_1$ numerically in time, i . e . , in going from j-th to
(j+1)-th step of iteration, 
say, $X_1$ has to be initialized as $X_1^{j0} = \frac{X_1^j}{d_j} d_0$. (The
time evolution of the other components of $X$ can be followed similarly).
This initialization implies that at each step, iteration starts with the same 
magnitude of $d_0$ but the direction of $d_0$ for step j+1 is that of $d(t)$
for j-th step (considered in terms of the ratio $\frac{X_1^j}{d_j}$).
For a more pictorial illustration we refer to Fig. 1 of Ref. \cite{ben}. 
j-th term of iteration means $t = j T$ $(j = 1, 2, \cdots \infty)$, where
$T$ is the characteristic time which corresponds to the shortest 
ensemble averaged period of nonlinear dynamical system.

Having obtained the time series in $X_1$ (it may be noted that the series in 
$X_i$-s are also required for calculation of the largest Lyapunov exponent as 
defined by 
\begin{eqnarray*}
\lambda = \lim_{
\begin{array}{c}
n \rightarrow \infty \\
d_0 \rightarrow 0
\end{array}
}
\frac{1}{nT} \sum_{j=1}^n \ln \frac{d_j}{d_0}
\end{eqnarray*}
for the chaotic system) the 
stationary probability density function $P_s(X_1)$ is computed as follows :
The $X_1$ axis ranging from -2 to +2 is divided into small intervals 
$\Delta X_1$ of size 0.025. The time series in $X_1$ is computed over
the time intervals
of 1000-10000 times of the time period $T$. For each time interval $\Delta X_1$
a counter is maintained and is initially set to zero before the simulation is 
started. 
The respective counter is incremented whenever $X_1$
falls within the given interval. Finally the steady state probability distribution
function $P_s(X_1)$ is obtained by normalizing the counts. The result 
is shown in Fig.1 for $\epsilon = 10$. Our numerical analysis shows that the
distribution function attains stationarity at around $t= 1000 T$, beyond which
no perceptible change in the distribution is obtained.

We now turn to the second issue. In what follows we shall be concerned with the
steady state entropy production (58) and its numerical verification. This
quantity can be calculated in two different ways. First, it may be noted that 
the determination of $S_{prod}$ (Eq.(57)) rests on two quantities defined in the 
tangent space; the steady state probability distribution function $P_s(U)$ and the 
diffusion coefficient ${\cal D}_s$ in $U$-space. Once the procedure for calculation of 
the distribution function as illustrated above is known from the time series in $X_1$ or $X_2$ the evaluation of 
$P_s(U)$  is quite  straightforward since $U$ is
expressed as $U = a_s X_1 + X_2$ according to (52). Here $a_s$ is given by (54)
with $\omega^2 = 2 b + c +c_2$ and the average $c$ and the integral over the 
correlation function, $c_2$ are as defined in (51). To calculate the correlation 
function $\langle \langle \zeta(t) \zeta(t-\tau) \rangle \rangle$ and the average
$\langle \zeta(t) \rangle$ it is necessary to determine long time series in $\zeta(t)$
($\zeta(t) = -12 a z_1^2$) by numerically solving the classical 
equation of motion (42) in phase space followed by averaging over the time series.
For further details of numerical analysis of correlation functions we 
refer to the earlier work \cite{sc3,bb1,bb2}.
The diffusion coefficient ${\cal D}_s$ can be numerically determined from the time series in $U$.
Having known $P_s(U)$ and ${\cal D}_s$ one can make use of the formula (57) to obtain the 
entropy production in the steady state. $S_{prod}$ is thus numerically 
calculated. The second procedure of calculation
of $S_{prod}$ is the direct theoretical evaluation of $\lambda_s$ from the expression (54).
Since the value of $\lambda_s$ rests again on $\omega^2$ and $\gamma$ and the
dependence of $\omega^2$ on the averages and  the correlation functions are
already known from the numerical 
analysis of phase space, $\lambda_s$ can be calculated
in the usual way. In Fig.2 we compare the values of steady state entropy production
thus obtained by the two different ways for several values of 
coupling-cum-field strength $\epsilon$. 
Here it should be noted that the curve connecting the squares (theoretically
calculated entropy production in the steady state) corresponds
to the negetive of entropy flux ($S_{flux}$) since for
the given example $S_{flux}$ in $U$ space is $-\lambda_s$ 
($-\int \lambda_s P_s(U) dU$, from Eq. (26)) for normalised
probability distribution function $P_S(U)$. Thus Fig.2 is a numerical proof
of $S_{prod} = -S_{flux}$.
The agreement is found to be quite
satisfactory. We therefore conclude that at least for the model studied here
and for the similar class of models the correspondence between the formulae
of steady state entropy production and the numerical computation is fairly
general.

\subsection{Entropy production in presence of weak forcing}

We now introduce an additional weak forcing in the dynamics. This is
achieved by subjecting the dissipative chaotic system to a weak magnetic
field ($\vec{B}$) through a velocity ($\vec{v}$) dependent force term
$\frac{e}{c_l} \vec{v} \times \vec{B}$
where $e$ and $c_l$ are the electric charge and
the velocity of light. For simplicity we apply the
constant field $B_z$ which is perpendicular to $q_1$-direction  [see Eq.39]. In
presence of this force field the motion of the particle will not be restricted 
to $B_z$ and $q_1$ only. We shall have to consider the other direction $q_2$
which is perpendicular to both $q_1$ and $B_z$.

To make the notation consistent with Eq.(8) we would now like to correspond
$X_1, X_2, X_3$ and $X_4$ to $\Delta q_1, \Delta q_2, \Delta p_1$ and 
$\Delta p_2$, respectively.

The relevant equations of motion are therefore as follows;
\begin{eqnarray}
\dot{X}_1 & = & X_3 \;, \nonumber \\
\dot{X}_2 & = & X_4 \;, \nonumber \\
\dot{X}_3 & = & 2bX_1-\gamma X_3+\zeta(t)X_1+
h \frac{e X_4}{c_l}B_z \; \;, \\
\dot{X}_4 & =& -h\gamma X_4 - h \frac{e X_3}{c_l} B_z \; \;. \nonumber
\end{eqnarray}
Here $e/c_l$ is the ratio of electric charge to the velocity of light used
to put the equation in appropriate dimension.

Then the non-equilibrium situation (due to additional forcing, $h\ne 0$)
corresponding to Eq.(59) is governed by
\begin{eqnarray}
\frac{\partial P}{\partial t} & = &
-h\frac{\partial}{\partial X_2}(X_4 P) +
h \frac{\partial}{\partial X_4}
(\gamma X_4)P-hB'_z \frac{\partial}{\partial X_3}(X_4P) \nonumber\\ 
& & +
hB'_z\frac{\partial}{\partial X_4}(X_4P)+{\cal D}_s \frac{\partial}{\partial U}
\left(P_s\frac{\partial}{\partial U}\frac{P}{P_s}\right)
\end{eqnarray}
where
\begin{eqnarray*}
B'_z & = & \frac{e}{c_l}B_z \\
{\cal D}_s & = & {\cal D}_{22}+{\cal D}_{21}a_s
\end{eqnarray*}
or more explicitly,
\begin{eqnarray}
\frac{\partial P}{\partial t} & = & -X_3 \frac{\partial P}{\partial X_1}
-\omega^2 X_1 \frac{\partial P}{\partial X_3}
+\gamma\frac{\partial}{\partial X_3}(X_3P) 
+h\gamma\frac{\partial}{\partial X_4}(X_4P)
-h X_4 \frac{\partial P}{\partial X_2}
\nonumber \\
& & - hB'_z\frac{\partial}{\partial X_3}(X_4P)+hB'_z\frac{\partial}
{\partial X_4}(X_3P)+{\cal D}_{21}\frac{\partial^2 P}{\partial X_3 \partial X_1}+
{\cal D}_{22}\frac{\partial^2 P}{\partial X_3^2} \; \;.
\end{eqnarray}

Proceeding as before we make use of the transformation of variables using
\begin{equation}
U'=a'X_1+b'X_2+c'X_3+X_4
\end{equation}
so that Eq.(61) becomes
\begin{equation}
\frac{\partial P}{\partial t}=\frac{\partial}{\partial U'}(\lambda' U'P)+
{\cal D}\frac{\partial^2 P}{\partial {U'}^2}
\end{equation}

\noindent
where
\begin{equation}
{\cal D}={\cal D}_{33}c'^2+ {\cal D}_{31}a'c' \; \;.
\end{equation}

\noindent
${\cal D}_{33}$ and ${\cal D}_{31}$ in four dimension correspond 
to ${\cal D}_{22}$ 
and ${\cal D}_{21}$ in two dimension, respectively and

\begin{equation}
\lambda' U'=-a'X_3-c'\omega^2X_1+\gamma X_3 c'+h\gamma X_4
-hB'_zX_4c'+hB'_zX_3 -hb'X_4 \; \;.
\end{equation}
Using (62) in (65) and comparing the coefficients of $X_i$-s we obtain
\begin{eqnarray}
b'=0 \; \; \; \; \; && a'=c'\gamma+hB'_z-\lambda' c' \nonumber \\
\; \; \; \; \; && c'=(h\gamma-\lambda')/hB'_z
\end{eqnarray}
where $\lambda'$ is a solution of the cubic algebraic equation
\begin{equation}
{\lambda'}^3+\lambda'(h\gamma^2+h^2{B'_z}^2-\omega^2)
-{\lambda'}^2\gamma(1+h)+\omega^2\gamma h =0 \; \;.
\end{equation}

We now seek for a perturbative solution of the algebraic equation (67)
which is given by (h as a small parameter) ;
\begin{equation}
\lambda'=\lambda_0'+\frac
{h\left({\lambda'}_0^2\gamma-\omega^2\gamma-\lambda'_0\gamma^2
-{\lambda'}_0 h {B'}_z^2\right)}
{3{\lambda'}_0^2-4\lambda'_0\gamma+h\gamma^2+h^2{B'}_z^2-\omega^2}
\end{equation}
where $\lambda'_0$ is the solution of (67) for $h=0$;
\begin{equation}
\lambda'_0=\frac{\gamma+\sqrt{\gamma^2+4\omega^2}}{2} \; \;.
\end{equation}
This is identical to $\lambda_s$ (Eq.54).
Therefore by virtue of Eqs.(66-69) all the constants in (62), 
i.e., $a', b', c'$ are now known. The stationary solution of (63) is now given by,
\begin{equation}
P'_s=N'e^{-\frac{\lambda' {U'}^2}{2{\cal D}}} \; \; ,
\end{equation}
where $N'$ is the normalization constant.

We are now in a position to calculate the steady state entropy flux
$\Delta S_{flux}$ due to external forcing ($h \ne 0$) from Eq.(38)
\begin{equation}
\Delta S_{flux}=h^2\int dX\delta P \nabla . F_1+h^2 \int dX
\left(\sum_i F_{1i}\frac{d\ln P_s}{dX_i}\right) \delta P \; \;.
\end{equation}
where the components of $F_1$ can be identified as
\begin{eqnarray}
F_{11}=0 \; \; \; \; \; && F_{13}=B'_z X_4 \nonumber \\
F_{12}=X_4 \; \; \; \; \; && F_{14}=-B'_zX_3-\gamma X_4 \\
{\rm and} \; \; \; \nabla_X . F_1=-\gamma \;, && \nonumber
\end{eqnarray}
and $h \delta P=P'_s-P_s$ denotes the deviation from the initial equilibrium
state due to external forcing. For normalized probability functions $P'_s$
and $P_s$ the first integral in (71) vanishes.

Since $P_s$ is given by (55) with $U$ as defined in (52), the expression 
for (71) reduces to
\begin{eqnarray}
\Delta S_{prod} & = & -\Delta S_{flux} \nonumber \\
& = & h B'_z\frac{\lambda_s}{{\cal D}_s}\int X_4 \left(a_sX_1+X_3\right)dX P'_s \; \;.
\end{eqnarray}
We now use the following transformations of variables 
\begin{eqnarray}
u' & = & a'X_1+c'X_3+X_4 \; \; \; \; \; ({\rm since} \; \; b'=0)
\nonumber \\
v' & = & X_3 \nonumber \\
w' & = & X_4 \nonumber \\
{\rm and} \; \; dX_1 dX_3 dX_4  & = & a' du' dv' dw' \; \;.
\end{eqnarray}
to calculate the integrals,
\begin{eqnarray}
\int X_3 X_4 dX P'_s & = & \frac{{\cal D}}{2\lambda'|c'|} \nonumber\\
{\rm and} \; \; \; 
\int X_1 X_4 dX P'_s & = & \frac{{\cal D}}{2\lambda'a'}
\end{eqnarray}
which yield
\begin{eqnarray}
\Delta S_{prod} & = & -\Delta S_{flux} \nonumber \\
& = & h B'_z \frac{\lambda_s}{{\cal D}_s} \frac{{\cal D}}{2\lambda'}\left[\frac{1}{|c'|}+
\frac{|a_s|}{a'}\right] \; \;.
\end{eqnarray}

For numerical verification of the above theoretical result (76) one can 
calculate entropy production ($\Delta S_{prod}$) numerically in the steady
state in presence of weak forcing from Eq. (37) as in the previous subsection.
The Eq. (37) for the present example reduces to the following form in the
steady state,

\begin{eqnarray}
\Delta S_{prod} & = & {\cal D}_{33} \int \int \int P_s' 
(\frac{\partial}{\partial X_3} \ln \frac{P_s'}{P_s})^2 dX_1 dX_3 dX_4 \nonumber\\
& & + {\cal D}_{31} \int \int \int P_s' 
(\frac{\partial}{\partial X_3} \ln \frac{P_s'}{P_s}) 
(\frac{\partial}{\partial X_1} \ln \frac{P_s'}{P_s})
dX_1 dX_3 dX_4 \; \;.
\end{eqnarray}

To calculate numerically $\Delta S_{prod}$ $\; \; \;$ 
${\cal D}_{33}, {\cal D}_{31}$
and $P_s$ can be determined by directly using the procedure
mentioned in the subsection VB and by simultaneously solving Eqs. (41) and (42).
Similarly one can calculate $P_s'$ from Eqs. (42) and (59). Finally
making use of all these quantities in Eq. (77) $\Delta S_{prod}$ can be obtained.
Thus the numerically evaluated $\Delta S_{prod}$ should correspond to the
results of Eq. (76) since our numerical verification in Fig.2 shows good 
agreement between numerical and theoretical results,  ${\cal D}_{33}, {\cal D}_{31}$
being very close to ${\cal D}_{22}$ and ${\cal D}_{21}$, respectively, since
$h$ is very small.

In the limit $h$ and $\gamma$ small the above expression (76) can be simplified
further. To this end we first note that
\begin{eqnarray}
|a_s| \sim \omega \; \; \; \; \; && \lambda_s \sim \omega
\; \; \; \; \; \lambda' \sim \lambda'_0 \sim \omega \nonumber \\
c'=\frac{a'}{\omega} \; \; \; \; \; && {\rm and} \; \; a'
=\frac{\omega h}{\gamma} B'_z \; \; .
\end{eqnarray}
This reduces ${\cal D}$ further as follows;
\begin{eqnarray}
{\cal D} & = & {\cal D}_{33}{c'}^2+{\cal D}_{31}a'c' \nonumber \\
& \simeq & \frac{{a'}^2}{\omega^2}({\cal D}_{33}-{\cal D}_{31}\omega) \; \;.
\end{eqnarray}
Thus we have
\begin{equation}
\frac{{\cal D}}{{\cal D}_s}=\frac{{a'}^2}{\omega^2} \; \;.
\end{equation}
Making use of (78-80), expression (76) can be approximated as
\begin{eqnarray}
\Delta S_{prod} & = & hB'_z\frac{{a'}^2}{\omega^2} \frac{1}{2}
\left(\frac{\omega}{a'}+\frac{\omega}{a'}\right) \nonumber \\
& = & \frac{h^2 e^2}{c_l^2 \gamma}B_z^2 \; \;.
\end{eqnarray}
This expression is due to average of the work per unit time of the external
force $B_z$ acting on the chaotic system. Note that the quadratic dependence 
on the magnetic field $B_z$ in Eq. (81) is characteristic of an expression
for entropy production in the steady state. Since the system is not 
thermostated this is independent of temperature. Although the leading order
expression (81) is apparently free from diffusion coefficients, a close look
into the more exact expression (76) reveals that their influence is quite 
significant in the higher order.

\section{Conclusions}
Ever since the development of the theory of chaos, the  
dynamical variables in the strong chaotic regime
have been interpreted as stochastic variables. One of
the earliest well-known examples in this connection was set by demonstrating \cite{casati} 
the linear
divergence of  mean-square momentum in time in standard map, mimicking
the Brownian motion. 
In the present paper we have tried to relate this chaotic diffusion to 
thermodynamic-like quantities by establishing a generalized Fokker-Planck
equation pertaining to the tangent space. The explicit 
dependence of drift and diffusion terms on the dynamical characteristics
of the phase space of the system is demonstrated.

The main conclusions of our study are;

\noindent
(i) We analyze the nature of chaotic diffusion in terms of the properties of the
phase space of chaotic systems. The drift and diffusion terms are dependent 
on the correlation of fluctuations of the linear stability matrix 
of the equation of motion.
Since the later is the key point for understanding the stability of
motion in a dynamical system, we emphasize that the thermodynamic-like 
quantities as discussed here have a deeper root in the intrinsic nature
of motion of a few-degree-of-freedom system.

\noindent
(ii) We identify the information entropy flux and production-like terms in the steady state
which explicitly reveal their connection to dynamics through
drift and diffusion term, in presence and absence of 
the external force field.

\noindent
(iii) The connection between the thermodynamically-inspired quantities and 
chaos are fairly general for the N-degree-of-freedom systems. 

The theory developed in this paper is based on the derivation of the
Fokker-Planck equation for chaotic systems pertaining to the processes
with correlation time which is short but finite (i. e. , for the systems
with hard chaos). The suitable generalization
of the approach to more general cases, where one encounters long correlation 
time is worthwhile for further investigation in this direction.

\acknowledgments
B. C. Bag is indebted to the Council of Scientific and
Industrial Research for a fellowship. 

\newpage
\begin{appendix}

\section{The derivation of the Fokker-Planck equation}

We first note that the
operator $\exp (-\tau \nabla_X \cdot {\bf L}^0)$ provides the solution 
of the equation [Eq.(\ref{e13}), $\alpha = 0$]
\begin{equation}
\frac{\partial f(X, t)}{\partial t} = -\nabla_X \cdot {\bf L}^0 f(X,t)
\label{a1}
\end{equation}
$f$ signifies the ``unperturbed'' part of $P$ which can be found 
explicitly in terms of characteristic curves. The equation
\begin{equation}
\dot X = {\bf L}^0 (X) 
\label{a2}
\end{equation}
determines for a fixed $t$  a mapping from $X(\tau=0)$ to $X(\tau)$, i. e., 
$X \rightarrow X^\tau$ with inverse $(X^\tau)^{-\tau}=X$ .
The solution of (\ref{a1}) is
\begin{equation}
f(X,t)= f(X^{-t}, 0) \left | \frac{d X^{-t}}{d X} \right | = 
\exp \left [ -t \nabla_X \cdot {\bf F}_0  \right ]  f(X, 0 ).
\label{a3}
\end{equation}
$\left | \frac{d(X^{-t})}{d(X)} \right |$ being a Jacobian determinant. The
effect of $\exp(-t \nabla_X \cdot {\bf L}^0)$ on $f(X)$ is as 
\begin{equation}
\exp(-t \nabla_X \cdot {\bf L}^0) f(X,0) = f(X^{-t}, 0) \left|\frac
{d X^{-t}}{d X} \right| \; \;.
\label{a4}
\end{equation}

This simplification in Eq.(\ref{e16}) yields
\begin{eqnarray}
\frac{\partial P}{\partial t} & = &
\left\{ -\nabla_X \cdot {\bf L}^0 
-\alpha \langle \nabla_X \cdot {\bf L}^1 \rangle + \alpha^2 \int_0^\infty
d\tau \left| \frac{d X^{-\tau}}{dX}
\right| \right. \nonumber \\
& & \langle \langle \nabla_X \cdot {\bf L}^1(X,t) 
{\bf \nabla}_{X^{-\tau}} \cdot {\bf L}^1({\bf x}^{-\tau}, t-\tau) \rangle 
\rangle \left. \left| \frac{dX}{d X^{-\tau}}
\right|  \right\} P \; \;.
\label{a5}
\end{eqnarray}

Now to express the Jacobian, ${X}^{-\tau}$ and $\nabla_{X^{-\tau}}$
in terms of $\nabla_X$ and $X$ we solve Eq.(\ref{a2}) 
for short time (this is
consistent with the assumption that the fluctuations are rapid \cite{van}). 
Using
Eqs.(\ref{e4}-\ref{e6}) we may rewrite ``unperturbed'' Eq.(\ref{a2}) as
\begin{eqnarray}
\frac{d}{dt}\left(
\begin{array}{c}
X_1\\
\vdots\\
X_N
\end{array}
\right) & = & \left(
\begin{array}{c}
X_{N+1}\\
\vdots  \\
X_{2N}
\end{array}
\right) 
\label{a6}\\
{\rm and} \; \; \; \; \; 
\frac{d}{dt}\left( \begin{array}{c}
X_{N+1} \\
\vdots \\
X_{2N} \end{array} \right) & = & -\gamma \left( \begin{array}{c}
X_{N+1} \\
\vdots   \\
X_{2N} \end{array} \right) + \left( \begin{array}{c}
G_{N+1}(X) \\
\vdots \\
G_{2N}(X)
\end{array} \right)
\label{a7}
\end{eqnarray}
Here $G_{N+1}(X) \cdots G_{2N}(X)$ are the functions of $\{X_i\}$ with
$i = 1 \cdots N$ only. This allows us to rewrite the solution of
(\ref{a6}) and (\ref{a7}) as,
\begin{equation}
\left(
\begin{array}{c}
X_1^{-\tau} \\
\vdots  \\
X_N^{-\tau}
\end{array}
\right)
= -\tau \left(
\begin{array}{c}
X_{N+1}\\
\vdots  \\
X_{2N}
\end{array}
\right)+\left(
\begin{array}{c}
X_1\\
\vdots  \\
X_N
\end{array}
\right)
=\left(
\begin{array}{c}
\bar{G}_1(X)\\
\vdots  \\
\bar{G}_N(X)
\end{array}
\right)
\label{a8}
\end{equation}
and
\begin{equation}
\left(
\begin{array}{c}
X_{N+1}^{-\tau} \\
\vdots  \\
X_{2N}^{-\tau}
\end{array}
\right)
= e^{\gamma \tau} \left(
\begin{array}{c}
X_{N+1}\\
\vdots  \\
X_{2N}
\end{array}
\right)-\tau\left(
\begin{array}{c}
G_{N+1}(X)\\
\vdots  \\
G_{2N}(X)
\end{array}
\right)
=\left(
\begin{array}{c}
\bar{G}_{N+1}(X)\\
\vdots  \\
\bar{G}_{2N}(X)
\end{array}
\right)
\label{a9}
\end{equation}
Here the terms of $O(\tau^2)$ are neglected. Since the 
vector $X^{-\tau}$ is expressible as a function of $X$ we write
\begin{equation}
X^{-\tau} = \bar{G}(X) \; \; ,
\label{a10}
\end{equation}
and the following simplification holds good;
\begin{eqnarray}
L^1(X^{-\tau}, t-\tau).\nabla_{X^{-\tau}} & = &
L^1(\bar{G}(X),t-\tau) . \nabla_{X^{-\tau}} \nonumber \\
& = & \sum_k L_k^1(\bar{G}(X),t-\tau)\frac{\partial}{\partial X_k^{-\tau}}
\nonumber \\
& = & \sum_j \sum_k L_k^1(\bar{G}(X),t-\tau)g_{jk}
\frac{\partial}{\partial X_j} \; \; \; \; \; 
\; \; \; ; j,k = 1 \cdots 2N
\label{a11}
\end{eqnarray}
where 
\begin{equation}
g_{jk}=\frac{\partial X_j}{\partial X_k^{-\tau}}
\label{a12}
\end{equation}

In view of Eqs.(\ref{a8}) and (\ref{a9}) we note:
\begin{eqnarray*}
{\rm if} \; \; \; j=k \; \; {\rm then} \; \; g_{jk} & = & 1, \; \; k=1 \cdots N \\
& = & e^{-\gamma \tau}, \; \; k=N+1 \cdots 2N\\
\\
{\rm if} \; \; \; j \ne k \; \; {\rm then} 
\; \; g_{jk} & \propto & -\tau e^{-\gamma \tau} \\
& {\rm or} & 0
\end{eqnarray*}
Thus $g_{jk}$ is a function of $\tau$ only.

Let 
\begin{equation}
R_j = \sum_k L_k^1(\bar{G}(X),t-\tau)g_{jk}
\label{a13}
\end{equation}
From Eqs.(\ref{e8}), (\ref{e9}) and (\ref{a10}) we write
\begin{equation}
L_i^1(X^{-\tau},t-\tau)
=L_i^1(\bar{G}(X),t-\tau) = 0 \; \; \; \; {\rm for} \; 
i=1 \cdots N
\label{a14}
\end{equation}
So the conditions (\ref{a13}), (\ref{a14}) and (\ref{a8}) imply that
\begin{eqnarray}
R_j(X, t-\tau) & = & R_j(X_1 \cdots X_N, t-\tau) 
\; \; \; {\rm for} \; j=1 \cdots N \nonumber \\
R_j(X, t-\tau) & = & R_j(X_1 \cdots X_{2N}, t-\tau) 
\; \; \; {\rm for} \; j=N+1 \cdots 2N
\label{a15}
\end{eqnarray}
We next carry out the following simplifications of 
$\alpha^2$-term in Eq.(\ref{a5}).
We make use of the relation (\ref{e10}) to obtain
\begin{eqnarray}
L^1(X,t).\nabla\sum_jR_j\frac{\partial}{\partial X_j} P(X,t)
& = & \sum_i L_i^1(X,t)\frac{\partial}{\partial X_i}\sum_j R_j
\frac{\partial}{\partial X_j}P(X,t)
\nonumber \\
& = & \sum_{i,j} L_i^1(X,t)
R_j \frac{\partial^2}{\partial X_i \partial X_j} P(X,t)\nonumber \\
&&+\sum_j
R'_j
\frac{\partial}{\partial X_j}P(X,t)
\label{a16}
\end{eqnarray}
where
\begin{equation}
R'_j=\sum_i L_i^1 (X,t) \frac{\partial}{\partial X_i} R_j
\label{a17}
\end{equation}
The conditions (\ref{a14}) and (\ref{a15}) imply that
\begin{eqnarray}
R'_j & = & 0 \; \; \; \; \; {\rm for} \; j=1 \cdots N \nonumber \\
R'_j & = & 
R'_j(X_1 \cdots X_N, t-\tau)  \ne 0 ;\; \; \; \; \; {\rm for} 
\; j=N+1 \cdots 2N 
\label{a18}
\end{eqnarray}
By (\ref{a18}) one has
\begin{equation}
R' . \nabla_X P(X,t) = \nabla_X . R' P(X,t)
\label{a19}
\end{equation}

Making use of Eqs.(\ref{e10}), (\ref{a11}), 
(\ref{a16}) and (\ref{a19}) in Eq.(\ref{a5}) we obtain the
Fokker-Planck equation (\ref{e17}).

\end{appendix}

\newpage
\begin{center}
{\bf Figure Captions}
\end{center}

\noindent
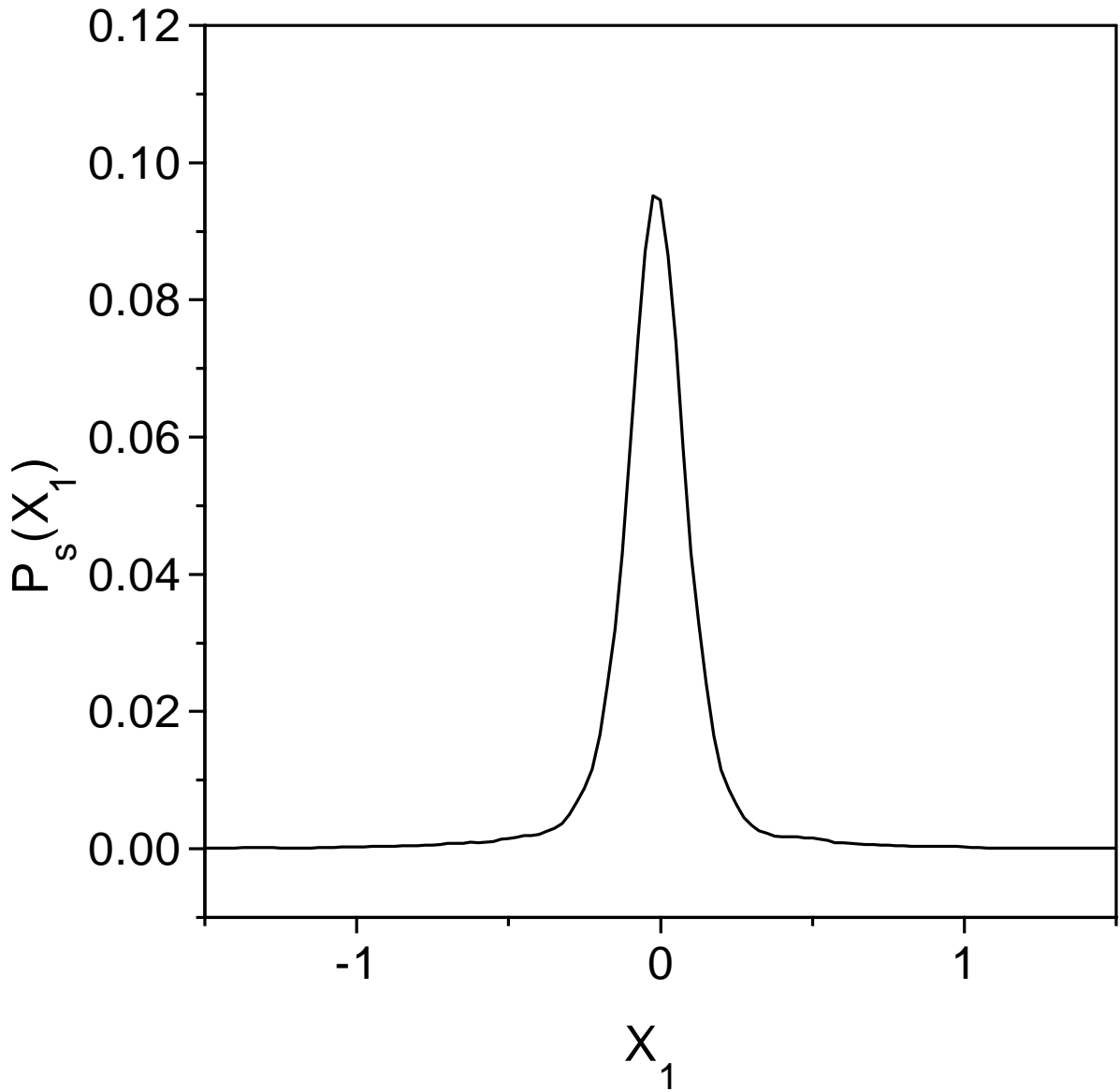
Fig.1. A plot of numerically calculated stationary distribution function 
$P_s(X_1)$ as a function of $X_1$ for the set of
parameter values described in Sec.V.

\noindent
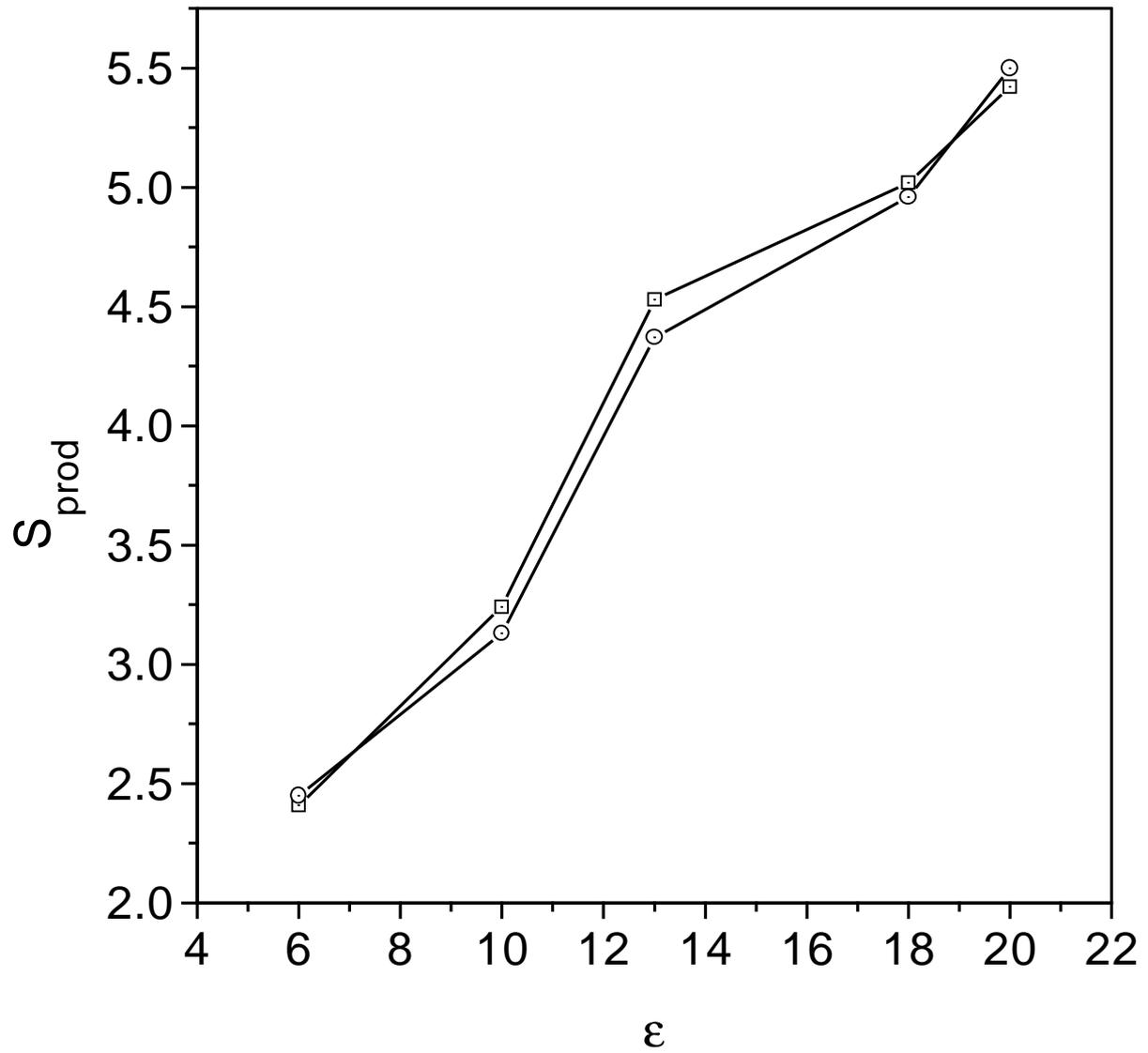
Fig.2. The steady entropy production calculated numerically (circle) 
and theoretically using Eq.(58) 
(square) for different values of driven field strength
$\epsilon$ for the model described in Sec.V.


\begin{thebibliography}{99}

\bibitem{kai}
T. Kai and K. Tomita, Prog. Theo. Phys. {\bf 64} 1532 (1980).

\bibitem{ono}
Y. Takahashi and Y. Oono, Prog. Theo. Phys. {\bf 71} 851 (1984).

\bibitem{pg}
P. Grassberger and I. Procaccia, Physica D {\bf 13} 34 (1984).

\bibitem{bonci}
L. Bonci, R. Roncaglia, B. J. West and P. Grigolini, Phys. Rev. Letts.
{\bf 67} 2593 (1993).

\bibitem{berdi}
V. L. Berdichevshy and M. V. Alberti, Phys. Rev. A {\bf 44} 858 (1991).

\bibitem{patt}
A. K. Pattanayak and P. Brumer, Phys. Rev. Letts. {\bf 79} 4131 (1997).

\bibitem{ruelle}
D. Ruelle, J. Stat. Phys. {\bf 85} 1 (1996).

\bibitem{cohen}
D. Cohen, Phys. Rev. Letts. {\bf 78} 2878 (1997).

\bibitem{gaspard}
P. Gaspard, J. Stat. Phys. {\bf 88} 1215 (1997).

\bibitem{fox}
R. F. Fox and T. C. Elston, Phys. Rev. E {\bf 49} 3683 (1994).

\bibitem{casati}
G. Casati, B. V. Chirikov, F. M. Izrailev and J. Ford in
{\it Stochastic behaviour in classical and Hamiltonian Systems}
(Lecture Notes in Physics, vol. 83) edited by G. Casati and J. Ford
(Springer, Berlin, 1979).

\bibitem{van}
N. G. van Kampen, Phys. Rep. {\bf 24} 171 (1976).

\bibitem{sc1}
S. Chaudhuri, G. Gangopadhyay and D. S. Ray, Phys. Rev. E {\bf 47} 311 (1993).

\bibitem{sc2}
S. Chaudhuri, G. Gangopadhyay and D. S. Ray, Phys. Rev. E {\bf 52} 2262 (1995).

\bibitem{sc3}
S. Chaudhuri, G. Gangopadhyay and D. S. Ray, Phys. Rev. E {\bf 54} 2359 (1996).

\bibitem{bb1}
B. C. Bag, S. Chaudhuri, J. Ray Chaudhuri and D. S. Ray, Physica D {\bf 125}
47 (1999)

\bibitem{bb2}
B. C. Bag and D. S. Ray, J. Stat. Phys. {\bf 96} 271 (1999). 

\bibitem{bb3} 
B. C. Bag and D. S. Ray, Phys. Rev. E {\bf 62} 1927 (2000).

\bibitem{zas}
G. M. Zaslavsky, Phys. Today, {\bf 52} 39 (1999).

\bibitem{sinai}
Ja. G. Sinai, Am. Math. Transl {\bf 31} 62 (1963) [See page 72].

\bibitem{Vasi}
See, for example, An Introduction to Statistical Physics by A. M. Vasilyev
(English translation, Mir publishers, Moscow 1983) page 53.

\bibitem{nicolis}
D. Daems and G. Nicolis, Phys. Rev. E {\bf 59} 4000 (1999).

\bibitem{jou}
A. Compte and D. Jou, J. Phys. A {\bf 29} 4321 (1996).

\bibitem{lin}
W. A. Lin and L. E. Ballentine, Phys. Rev. Letts. {\bf 65} 2927 (1990).

\bibitem{ben}
G.Benettin, L. Galgani and J. Strelcyn, Phys. Rev. A {\bf 14} 2338 (1976).

\end{thebibliography}
\end{document}